\documentclass[11pt]{report}
\usepackage{color}
\usepackage{graphicx}
\usepackage{harvard}

\begin{document}

\baselineskip=.60cm

\centerline{ \Large{\bf Note about the impact possibilities of
asteroid (99942) Apophis}}

\vspace{0.4cm}

\centerline{by}

\vspace{0.4cm}

\centerline{\bf Ma{\l}gorzata Kr{\'o}likowska and Grzegorz Sitarski}

\vspace{0.5cm}

\centerline{Space Research Centre, Polish Academy of Sciences,}
\centerline{Bartycka 18A, 00-716 Warsaw, Poland}

\centerline{e-mails: mkr@cbk.waw.pl; sitarski@cbk.waw.pl}

\vspace{1.2cm}

{\small \centerline{A B S T R A C T}

\vspace{0.4cm}

 The Monte Carlo method of the nominal orbit clonning was applied
to the case of 99942 Apophis, the asteroid from the Aten group.
Calculations based on observations from the time interval of
2004\,03\,15~-–~2008\,01\,09 have shown that the asteroid will pass
near Earth in 2029 at the minimum distance of $5.921\pm
0.042$~R$_{\rm Earth}$, what implies that the likelihood that
Apophis strikes the planet at 2036~April~13 increased to 4.5$\times
10^{-6}$ (from about 6$\times 10^{-7}$ previously announced by us in
Paper~I \cite{KSS09}). This value is identical with that given by
Chesley, Baer, and Monet \citeyear{Ches10}.}

\vspace{0.7cm}

\centerline{\bf 1. Introduction}

\vspace{0.4cm}

This note is motivated by the the article {\it Treatment of star
catalogue biases in asteroid astrometric observation} by Chesley,
Baer, and Monet \citeyear{Ches10}. That paper presents interesting
discussion about systematic errors in star positions of the USNO
series of star catalogs often used to reduce the astrometric
observations of currently observed asteroids in recent years. In the
abstract, the authors wrote that the inclusion of the proposed
debiasing techniques to improve the astrometric observations of the
asteroid 99942 Apophis reduces the impact probability  ''by nearly
an order of magnitude to $4.5\times10^{-6}$ for the 2036 close
approach''.

Earlier this year we released a new analysis of the asteroid Apophis
collision in our Web Page \cite{SRCpage} (in February 2010) and in
the Magazine of Polish Academy of Science \cite{KroSli09}. The first
source (website) mainly presents examples of the orbital elements of
the impact orbits for potentially dangerous future dates. However,
we would like now to show that our method of weighting the
observations seems to effectively eliminate problems associated with
the existence of systematic biases in some sequences of observations
for asteroid Apophis case.

\vspace{0.4cm}

\vfill\eject

 \centerline{\bf 2. The method}

\vspace{0.4cm}

Results presented here are based on the archive positional
observations taken from the NEODyS (Near Earth Object - Dynamic
Site, University of Pisa, Italy) available on the Web at
http://newton.dm.unipi.it/neodys/. The whole observational material
contained 1399 observations covering the time period from
March~15,~2004 to January~9, 2008. The data were taken from NEODyS
source at the end of 2009. However, the current observation interval
has not changed (September~14, 2010).

Our method of the data treatment as well as the full method of
impact orbit determination are described in details in Paper~I.
Therefore, current results differ from that paper only because of
longer arc of observations and much greater number of observations.
We derived the nominal orbit using this longer arc with
rms$=0.270$\,arcsec (2774 residuals).

\vspace{0.7cm}

\centerline{\bf 3. Results}

\vspace{0.4cm}

\begin{figure}
\begin{center}
\includegraphics[width=15.0cm]{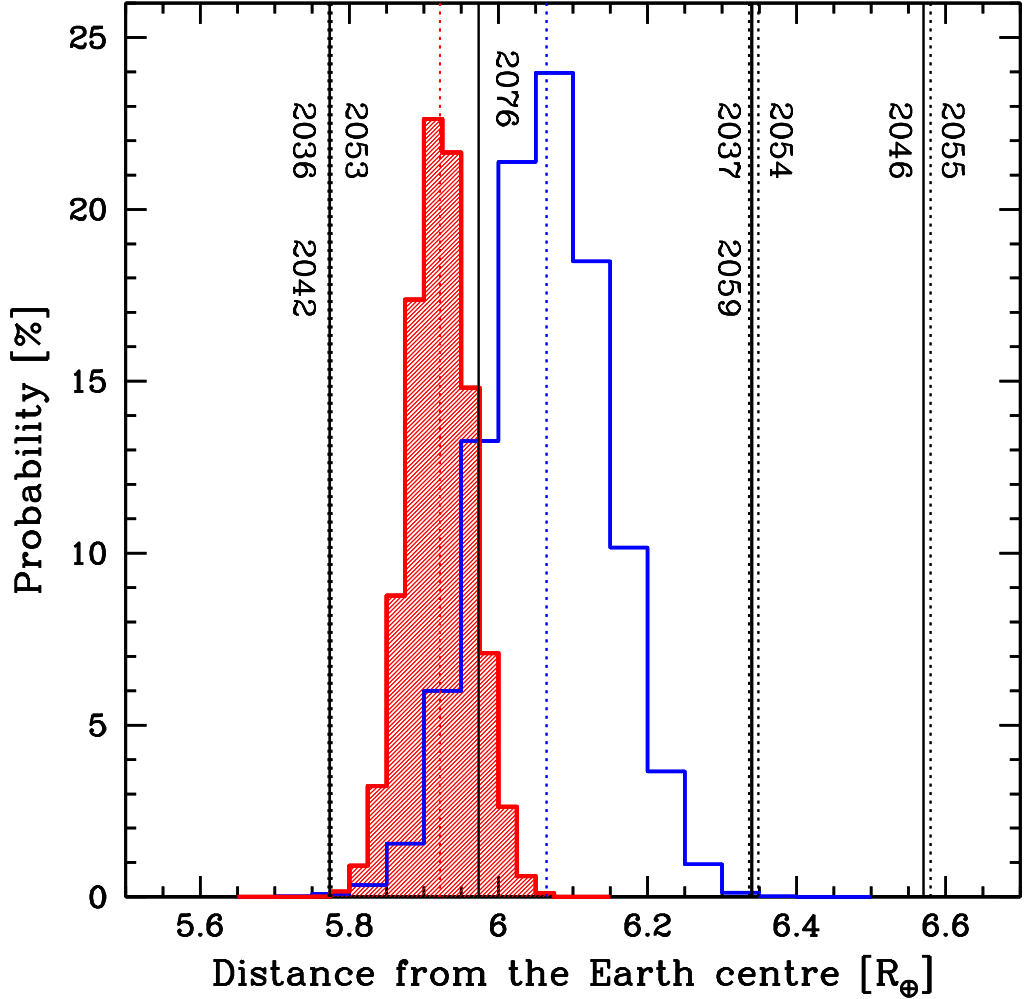}
\end{center}
\caption{Distributions of the minimum distance of the asteroid
Apophis from the centre of Earth in 2029 04 13 derived for the
samples of $15\,000$ virtual orbits.  The blue histogram shows
results based on a shorter observational arc (Model ~E from the
Paper~I), while for the current model based on longer arc is shown
with a red solid line and filled histogram. Thin dotted vertical
lines represent the position of nominal orbits derived for Model~E
and for current model and are shown with a blue and red ink,
respectively. Solid and dashed black vertical lines represent
gravitational keyholes, marked with dates of possible future
collisions. The possible collision in 2076 could only occur if
preceded by a flyby at a very precise distance in 2051. Thus, the
2029 keyhole for the impact in 2076 is extremely small and the
probability of such collision turns out to be significantly smaller
than the probability of impact in 2036. } \label{fig:EarthDist}
\end{figure}

The distribution of minimum asteroid distance from Earth is
presented for the current solution by magenta filled histogram in
Fig.~\ref{fig:EarthDist}. The minimum-distance histograms for
shorter arc (best Model~E from the Paper~I) is shown  with a blue
solid line.

We obtained 45 impact orbits from the sample of $10^7$ VAs (Virtual
Asteroids) for the Earth encounter on April 13, 2036 using the
methods described in details in Paper~I. Thus, the impact
probability of $\sim 4.5\cdot 10^{-6}$ was calculated for the
current solution. The best solution in the Paper I (Model E) gave
the probability of a collision in 2036 at the level of $\sim 6\times
10 ^{-7}$, which is almost an order of magnitude smaller than the
current solution based on the longer arc of observations. This is
explained in Fig.~1. Figure~1 also shows that the impact
possibilities in 2037 and years related to the close encounter in
2037 should now be excluded because of practically zero likelihood
of such events (while in Model E from the Paper I the impact
probability in 2037 was about four times greater than probability
for the year 2036).

The coincidence of our impact probability value for a potential
collision in 2036 for the Apophis case with the value obtained by
Chesley, Baer, and Monet \citeyear{Ches10} indicates that our
weighting procedure of observations seems to effectively tackle
problems associated with the existence of systematic biases in some
sequences of observations.

\vspace{0.3cm}

\end{document}